# Multimodal near-infrared spectroscopy for biosensing applications

Krisztian Neutsch[1], Jana Nikolić[1], Asra Mafakheri[3], Jan Stegemann[1,2], Sebastian Kruss[1,2]*

[1] Department of Chemistry and Biochemistry, Bochum University, Bochum, Germany
[2] Fraunhofer Institute for Microelectronic Circuits and Systems, Duisburg, Germany
[3] Department of Physics, Institute for Advanced Studies in Basic Sciences, Zanjan, Iran

* Email: Sebastian.kruss@rub.de

**Abstract**

The near infrared (NIR) region of the electromagnetic spectrum contains valuable information about vibrational overtone and combination modes. Additionally, fluorophores that emit in the NIR are desired because this spectral range falls into the tissue transparency window. However, NIR techniques are not as widespread as their counterparts in the visible range because detectors, microscopes and fluorophores are less available. Here, we present a NIR multimodal and low footprint spectroscopy (NIR-MMS) setup. The setup excites NIR fluorescence using a LED or uses broadband NIR light to capture absorption spectra in the range from 900 nm to 1650 nm with a fiber-coupled InGaAs spectrograph. We demonstrate the measurement of NIR fluorescent materials and biosensors for the neurotransmitter dopamine. Additionally, the setup allows to simultaneously measure absorption and emission spectra to assess chemical changes in NIR fluorescent carbon nanotubes. Overall, this small footprint setup provides simultaneous access to NIR fluorescence and absorption spectra with broad applications in biosensing.

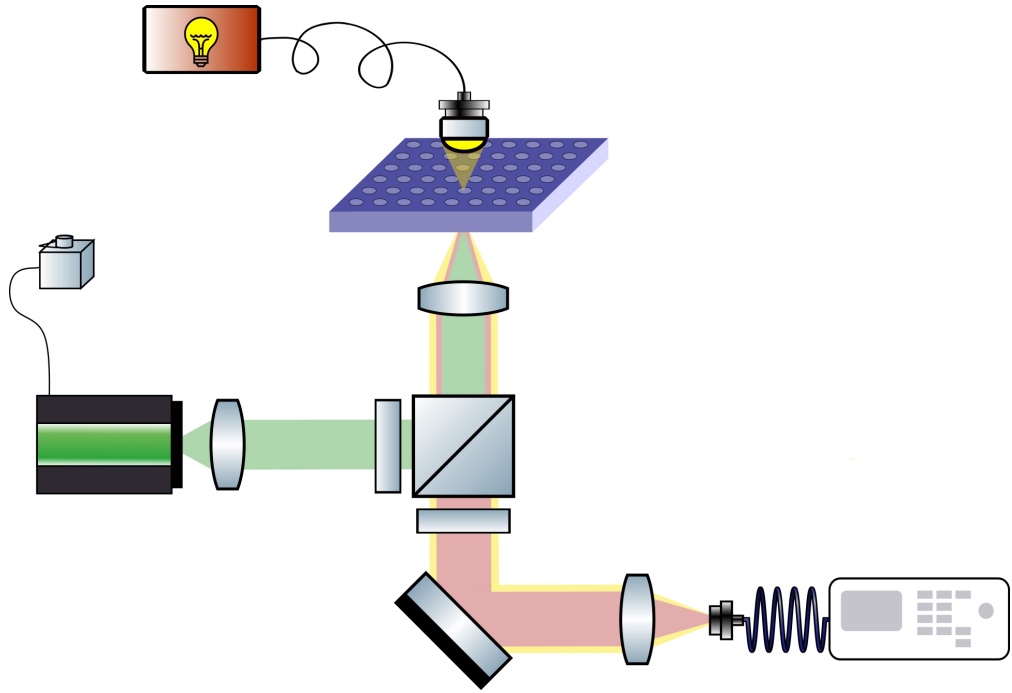

**TOC figure**

## Introduction

Spectroscopy plays a foundational role in all scientific disciplines from chemistry, physics, biology, materials science, environmental science to astronomy[1]. Portable and handheld spectrometers enable on-site analysis, thus facilitating rapid, informed decision-making without the need to send samples to a laboratory[2].

The complexity of many materials and biological samples demands an analytical approach that can extract both qualitative and quantitative information. Near infrared (NIR) spectroscopy satisfies many of these requirements and is a powerful and widely utilized techniques. It enables non-invasive characterization of materials and complex biological samples[3].

The NIR spectral region contains mainly overtones, combination modes, and resonances of the fundamental molecular vibrations[4,5]. These NIR-active vibrational modes are predominantly associated with the highly anharmonic modes of chemical groups containing relatively heavy atoms (C, N, O, and S) bonded to hydrogen atoms [6].

On the other hand, the NIR-I (650 nm – 900 nm) and NIR-II (1000 nm – 1700 nm) regions overcome limitations of biomedical sensing in the visible (400 nm – 700 nm) spectrum[5,7]. There are lower autofluorescence levels from organic molecules and less scattering, resulting in a reduced background signal and enhanced contrast[8]. These two optical transparency windows provide benefits for photonics by enabling deeper tissue penetration, higher spatiotemporal resolution, and lower phototoxicity[9]. The detection of NIR signals is challenging, since the sensitivity of standard Si-based detectors is restricted to 1100 nm. Therefore, InGaAs-based detectors are typically employed for NIR signal in the range 900 to 1700 nm. Compared to Si, these detectors are more expensive, typically need cooling due to higher dark currents and possess a lower amount of pixels, limiting the development of NIR signal-based biosensors and devices[10,11].

The beneficial properties of the NIR accelerated the development of NIR fluorescent biosensors. Such sensors leverage the remarkable sensitivity and selectivity of biological systems combined with physicochemical transducers to enable bioanalytical measurements in user-friendly platforms[12]. In this context, various NIR-II fluorophores such as rare earth-doped nanoparticles, quantum dots, small molecule dyes, gold nanoclusters, and single-walled carbon nanotubes (SWCNTs) are used. These fluorophores offer deep tissue penetration, minimal background interference, and high-resolution[13], for applications including bioimaging, drug delivery[14], biosensing, molecular and cellular level studies, and clinical translation[15]. Upon excitation of SWCNTs with visible light, an electron transition from valence to conduction band generates an electron-hole pair (exciton)[16]. Typically, for semiconducting SWCNTs absorption with visible light excites an electron to the second conduction band, from where it relaxes to the first conduction band from which the exciton can recombine radiatively, emitting NIR fluorescence as a result[17]. The exciton is highly sensitive to the chemical surrounding, making SWCNTs versatile and modular building blocks, enabling a wide range of applications[17,18]. Notably, they are widely used in in vivo imaging applications, providing deep tissue penetration and high spatial resolution[5,19]. Furthermore, the wavelength and intensity of SWCNTs emission can be significantly altered through surface functionalization or interaction with target analytes, rendering them well-suited for fluorescence-based sensing and detection[17] of different molecules such as pH[20], neurotransmitters[21], proteins[22], mRNAs[23], signaling molecules[24], hydrogen peroxide[25,26,27], and others[28,17]. Another fluorophore that fluoresces in the NIR range is the ancient inorganic pigment known as Egyptian Blue (EB), which offers optimal properties for photonics and bioimaging applications[29]. It can be easily exfoliated into two-dimensional nanosheets that preserve the photophysical characteristics of the bulk material[30].

In the field of biosensing, fluorescence spectroscopy is a highly efficient technique for detecting biomolecules[31]. Absorption spectroscopy, on the other hand, involves measuring the amount of light absorbed by a sample as a function of the wavelength. This technique is widely used in environmental analysis for the determination of molecular species[32]. Fluorescence spectroscopy enables the detection of SWCNTs based on their unique photoluminescence properties in the NIR region[17,33]. Conversely, absorbance spectroscopy provides valuable insights into the electronic transitions and structural characteristics of SWCNTs, allowing for a deeper understanding of their optical properties [34]. When utilizing SWCNTs, the integration of both fluorescence and absorbance measurements in a small footprint single device would allow a comprehensive analysis of samples, providing complementary information about their optical properties.

Here, we present a portable, low-cost setup capable of performing absorption/reflection and fluorescence spectroscopy in the NIR region from 900 to 1650 nm. This multimodal setup enables versatile characterization of nanomaterial-based biosensors such as NIR fluorescent SWCNTs and Egyptian Blue nanosheets (EB-NS). Additionally, photophysical effects related to aggregation or the formation of quantum defects can be monitored. The portability and low-cost design promises applications for example in biosafety labs without optical tables and bulky specialized optical setups. This design of the optical setup closes a gap that hindered research groups from exploring the NIR region for characterization and applications of fluorescent nanomaterials.

**Results**

**Design of the NIR-MMS setup**

The NIR-MMS setup is based on a simple optical configuration to enable the two measurement modes (Figure 1). For the fluorescence spectrometer mode, a green LED ($\lambda$ = 565 nm, $P_{typ}$ = 979 mW, Thorlabs M565L3) is employed and collimated by lens L1 (f = 20.1 mm, Thorlabs SM1U25-A). The collimated light is directed onto the filter cube, which contains a shortpass (SP) filter for suppressing NIR radiation of the illumination source ($\lambda_{Cut-off}$ = 650 nm), a dichroic mirror with $\lambda_{Cut-on}$ = 805 nm for separation of the excitation and emission of fluorophores, as well as a long-pass filter ($\lambda_{Cut-on}$ = 850 nm) to suppress remaining visible radiation leaking into the detector arm. The filter cube and the LED are exchangeable, which allows for applications with different fluorophores with varying fluorescent spectra.

The dichroic mirror directs the green excitation light towards the sample plane. Here, a microscope objective (L2, Olympus LUCPlanFL N, 20x, NA = 0.45) focuses the light into a single well of a 96-well plate, which is placed on a motorized XY stage (Zaber ASR100B120B-T3). For the sake of simplicity, one could replace this stage by a petri dish holder or any other sample holder. However, the stage allows fast scanning of entire well plates automatically and is thus suitable e.g. for high throughput screenings of nanosensors[35].

The NIR fluorescence signals are captured by the objective and directed back to the dichroic mirror. Since the signal is above the cut-on wavelength of both the dichroic mirror and the shortpass, it is transmitted into the detector arm. Lens L3 (N-BK7 Plano-Convex, f = 50 mm) narrows the beam diameter of the NIR signal to match the aperture of the fiber-collimator of the detector. After the collimator, a fiber bundle (PL100-2-VIS-NIR) transmits the signal to the InGaAs detector (Ocean Insight, NIRQUEST+1.7, 512 px, 900 - 1650 nm, -20° TE-cooling). The detector's spectral resolution is 2.8 nm, and the readout of the signal takes place via USB for example with a laptop computer.

The absorption spectroscopy mode follows the exact same detection path, but a different illumination path. Instead of the LED illumination from the bottom, a broadband lamp (Stellar-Net SL-1, 350 - 2500 nm) is used in transmission. It is connected to a fiber and a fiber-collimator (Stellar-Net F600-VISNIR and LENS-COL) and illuminates the sample in transmission mode from the top. The light transmitted through the sample is captured by the objective and directed into the detector in the same manner as before.

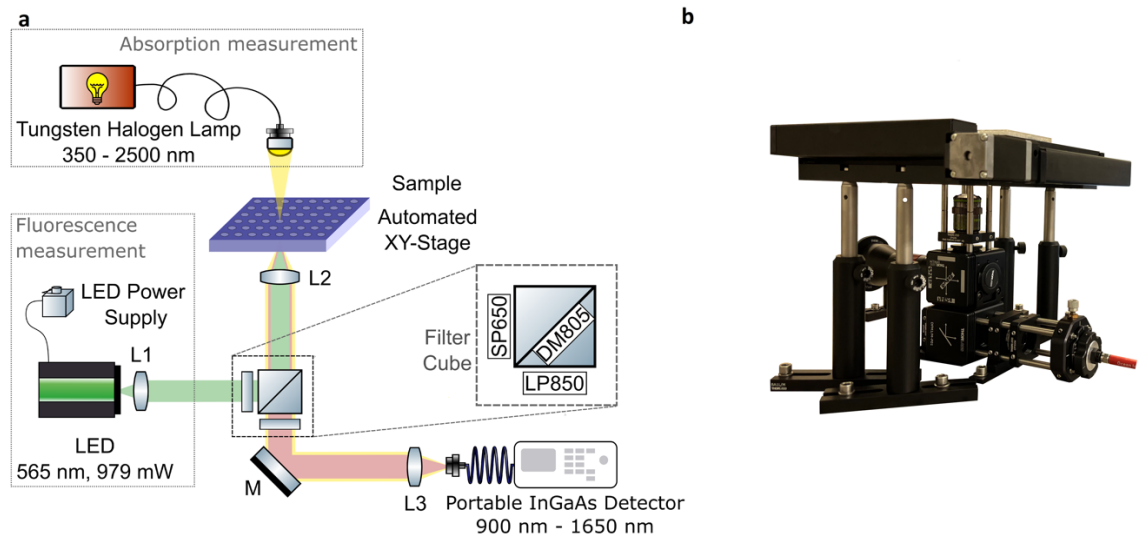

*Figure 1. Design of the multimodal NIR spectroscopy (NIR-MMS) setup. a) Schematic of the setup with two illumination modes for absorption and fluorescence measurements and a fiber-coupled small footprint InGaAs spectrometer. b) Picture of the setup. Note that the size of the mechanical components could be further minimized and the by far largest component is the xy stage. L1, L2, L3: Lenses, M: Mirror, SP: Shortpass, DM: Dichroic Mirror, LP: Longpass.*

A detailed list of components and their price is provided in the SI. Main parameters for the measurement are the integration time of the spectrometer, the number of scans to average the results and choice of the measurement mode. Measurement results are processed only with a gliding average filter to reduce noise. In the fluorescent measurement mode, the results are used only in a wavelength range until 1300 nm, which is the range of interest for the fluorophores used in our study. The detector itself can capture until 1650 nm, but one has to make sure that the dichroic filters still transmit efficiently. In absorption measurements, the limit does not exist, as the filter cube is removable and not necessary for these measurements. However, for fast quasi-simultaneous measurements the filter cube is required.

**Characterization of NIR fluorescent samples**

To demonstrate the possible applications of the NIR-MMS setup, experiments with different materials/fluorophores were performed. A focus were scenarios in which the optical properties change, and simultaneous collection of absorption/fluorescence spectra adds important information and value. For the wavelength range between 900 nm and 1300 nm we used SWCNTs functionalized with either ssDNA or sodium dodecyl benzenesulfonate (SDBS), EB-NS, and the organic dye IR-1061.

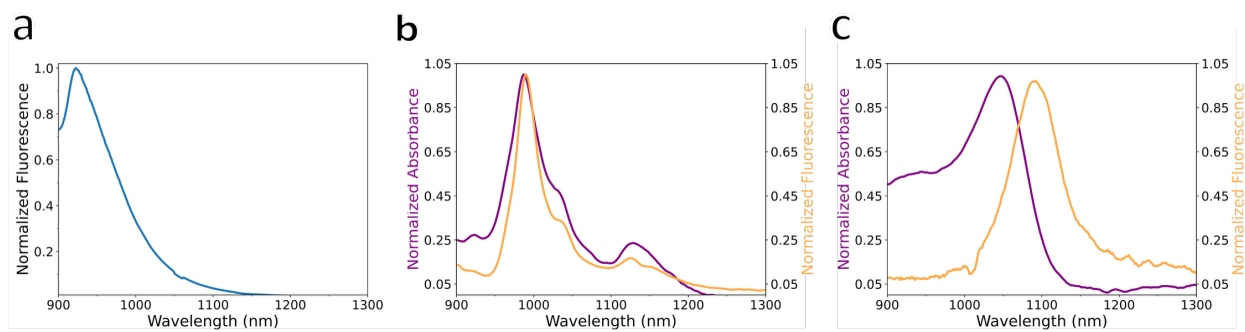

*Figure 2. Fluorescence and absorption spectra of NIR fluorophores.* a) Fluorescence spectrum of calcium copper silicate nanosheets (EB-NS) in water. b) Fluorescence and absorption spectra of 10 nM $(GT)_{10}$-SWCNTs in PBS. c) Normalized absorbance and fluorescence spectra of the NIR fluorescent organic dye IR-1061 (100 µg/ml in ethanol).

EB-NS have attracted attention as NIR fluorophores due to their outstanding photoluminescence properties and many possible applications[36]. The absorption spectrum of EB-NS in water is dominated by scattering, however their fluorescence spectra show an intense peak at around 920 nm (Figure 2a). This spectrum is most likely caused by the $B_{2g} \rightarrow B_{1g}$ transition of the $Cu^{2+}$ ion[37,38]. The luminescence of SWCNTs originates from the so-called $E_{11}$ transition at the bandgap of these semiconductors. The SWCNTs used in our sample contained mainly the (6,5)-chirality and can be excited with green light at the $E_{22}$ transition. The $E_{11}$ absorption is also observable in the absorption spectrum. Both absorption and emission spectra can be measured by our NIR setup and show the small Stokes shift (2.7 nm) between $E_{11}$ emission and absorption (Figure 2b).

Next, the absorbance and fluorescence spectrum of a typical NIR organic dye (IR-1061) was measured (Figure 2c). The wavelength maximum in the absorbance spectrum appeared at 1046 nm, while the emission showed a maximum at 1090 nm. The absorption spectrum indicates that an ideal excitation wavelength is above 800 nm. Even though our setup so far only uses a single LED light source (565 nm) for the fluorescence measurements, the emission spectrum of the dye remains unchanged because of quick vibrational relaxation (ps time scale) into the ground vibrational level of excited electronic state from which fluorescence emission occurs[39]. The IR-1061 dye is only soluble in various organic solvents. It has been shown that IR-1061 dye can dimerize[40,41], which creates a broad absorption peak at around 900 nm. The broad peak at around 900 nm (Figure 2c) showed the presence of this dimer under our experimental conditions and the monomer absorbed at around 1090 nm.

**Measurements of NIR fluorescence changes for biosensing**

In the presence of analytes such as the important neurotransmitter dopamine[42] the fluorescence intensity of SWCNTs can change. Selectivity is created by chemical tailoring and as discussed above there are chemical designs for a huge range of analytes[17]. This is the basis for using SWCNTs as biosensors. NIR-MMS setup can capture biosensor responses such as the concentration dependent response of $(GT)_{10}$-SWCNTs to dopamine (Figure 3a).

The mechanism of this fluorescence change has been linked to the interaction between dopamine and ss-DNA. More specifically, the two OH groups of dopamine interact with phosphate groups of the DNA wrapped around SWCNT[21,43,44]. These conformational changes affect the local solvation, which changes the exciton decay pathways[45].

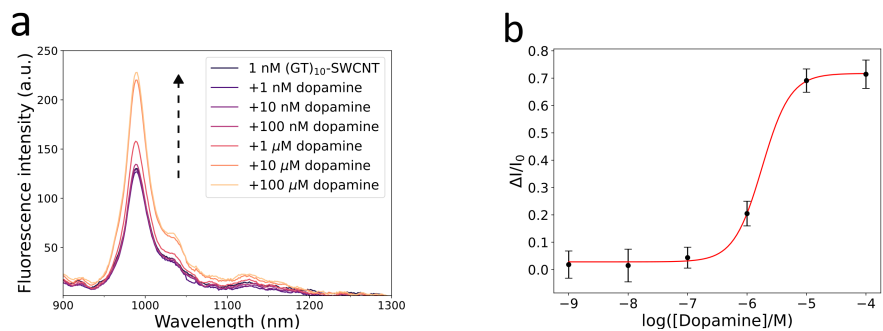

*Figure 3. Measurement of NIR fluorescent biosensor responses. a) Fluorescence spectra of 1 nM $(GT)_{10}$-SWCNTs in PBS before and after adding different dopamine concentrations. b) Dose response curve (red= fit, see equation 1 below, $R^2$ =0.99809) of $(GT)_{10}$–SWCNTs in PBS upon adding dopamine. Data = mean ± SD, n=3.*

$(GT)_{10}$-SWCNTs responded by a signal increase in the presence of analyte concentrations ≥ 10 nM, which our setup can detect. The maximum intensity wavelength did not shift to lower or higher wavelength, but remained at approximately 990 nm, which corresponds to the $E_{11}$ transition of the (6,5)-SWCNT chirality[46]. The response of $(GT)_{10}$ –SWCNTs to dopamine was modeled using a sigmoidal dose-response curve:

$$y = \frac{A}{1 + 10^{(\log_{10}(EC50)-x)p}} \qquad 1$$

In equation **1**, parameter A represents upper limit of the response, logEC50 is the dopamine concentration that produces a half-maximal response, and p is the variable Hill slope. From the performed fit, the Hill slope was found to be p = 1.8 ± 0.4, and the dopamine concentration at the half-maximal response (EC50) was (1.8 ± 0.3) µM. These numbers are similar to reported values of dopamine sensitivity and dynamic range[21,43,44] but were acquired with our low-footprint dual absorption/fluorescence setup.

Next, the chemical reaction of SDBS dispersed SWCNTs with diazonium salts under constant LED illumination was investigated. This reaction introduces so called $sp^3$ quantum defects on the SWCNT surface, which creates a red-shifted $E^*_{11}$ emission. The challenge is to create sufficient quantum defects that serve as exciton traps but not too many because they would destroy the whole NIR fluorescence of the SWCNTs[47]. Therefore, the suitable concentration of diazonium salt, reaction times and conditions have to be identified[49,50,51]. For our experimental conditions, it was determined that the optimal concentration of the salt lies in the region between 1 and 2 µM. The presence of a new $E^*_{11}$ band at around 1136 nm upon adding 2 µM diazonium salt indicated the introduction of defects (Figure 4a). The $E_{11}$ band maximum was observed at 977 nm and its intensity decreased over time.

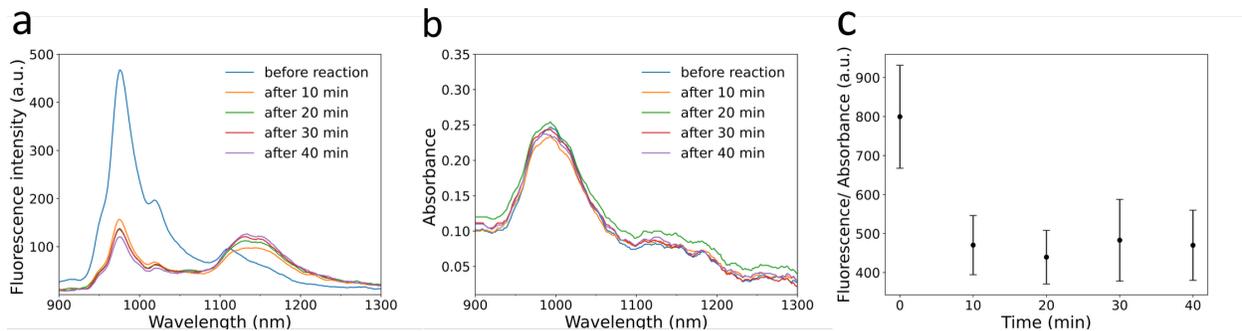

*Figure 4. Parallel absorption and fluorescence spectroscopy to follow the formation of quantum defects in SWCNTs. Fluorescence (a) and absorbance spectra (b) of 0.5% SDBS dispersed 5 nM SWCNTs before and after adding 2 µM diazonium salt. (c) Ratio of the total fluorescence and absorbance area over the course of the reaction, which corresponds to the relative quantum yield. Data = mean ± SD, n=3.*

A major advantage of our NIR-MMS setup is the ability to quickly switch between absorbance and fluorescence measurements. After capturing background spectra, one can place the sample in the well plate and first collect absorbance spectrum, then switch to fluorescence mode and capture the emission signal. After adding 2 µM diazonium salt and continually illuminating the sample with 565 nm LED, spectral data at different time periods were captured. During time intervals of 10, 20, 30 and 40 minutes, the LED was quickly turned off and absorbance mode was selected to record absorbance spectra, after which the LED was again turned on, so that the reaction could continue. Consequently, quasi-simultaneous fluorescence and absorbance spectra of 5 nM SWCNTs dispersed in 0.5% SDBS after adding 2 µM diazonium salt could be measured. It should be noted that the choice of the surfactant not only affects the properties of spectra, but also the incorporation of defects onto the SWCNT surface. Less rigid, flexible surfactants such as SDS and SDBS allow diazonium to covalently bind to the SWCNT, resulting in significant quenching of the fluorescence $E_{11}$ peak. In contrast, when used as surfactants, bile salts such as DOC and SC, are much more rigid and pack more densely around the nanotube, thereby hindering the accessibility of diazonium to the SWCNT surface and slowing down the reaction[52]. Both absorption and fluorescence spectra contain important information, and one question is how the quantum yield of SWCNTs changes when quantum defects are incorporated. The quantum yield is defined as the ratio of number of emitted photons to absorbed photons. Absolute quantum yield measurements can be performed with integrating sphere setups. However, the number of emitted photons is proportional to the area under the fluorescence curve, and absorbed photons are proportional to the integrated optical density of the sample[53]. Hence, the ratio of the fluorescence and absorbance area provides a relative estimate of the quantum yield of the sample under investigation. Quantum yield is one of the most important properties of a sample, and can change due to various factors, such as environmental and solvent effects[54]. With our setup it was possible to follow the change of the relative quantum yield (Figure 4c). The data showed that the absorption spectra only slightly changed while the overall fluorescence immediately decreased. It can be attributed to a direct quenching effect of the diazonium salts that are present in the sample. After dialysis it is expected that the overall quantum yield of the SWCNTs is higher[47]. Overall, one can therefore optimize reaction conditions not only for one readout but use this information to optimize the quantum yields. The NIR-MMS setup shares similarities with budget-friendly custom built NIR fluorimeters e.g. by Kallmyer et al[10]. However, our setup uses fiber coupled thermoelectrically cooled InGaAs array detector, which comes at a higher cost but provides important spectral information instead of fluorescence intensity only. Moreover, our setup can perform fluorescence and absorbance measurements simultaneously, enhancing its versatility.

**Conclusion**

We present a simple portable device (NIR-MMS) for multimodal spectroscopy in the NIR, which detects absorption and fluorescence of NIR fluorophores and specifically nanosensors. It allows their rapid investigation without the need for an optical table or a large-footprint spectrometer. Its portability allows its usage in labs with biosafety level (BSL) or point of care (POC) diagnostics, where samples cannot be transported to optical labs for investigation. This is especially relevant in settings as for example in precision agriculture that so far only used single pixel detectors for NIR measurements of SWCNT-based biosensors[10,55].

For fluorescence measurements, the LED illumination and filter cube is adaptable to the fluorophore's excitation spectrum. For absorption measurements, a broadband lamp allows for transmission illumination. Signals from both measurements are captured with a cooled InGaAs detector. Further noise reduction is provided by sampling multiple scans, averaging and numerical filtering. This setup could be further modified to contain a laser for excitation and a Vis fiber coupled spectrograph to collect as well Raman spectra. This way one could obtain the full optical information from most samples.

We showcase the potential of the setup to study static as well as dynamic changes in NIR signals. Therefore, it is useful for biosensing or reaction monitoring. The main advantage is that it can measure both absorption and fluorescence spectra. Additionally, the xy stage provides the opportunity for automated measurements of samples in well-plates, which is a key challenge for high throughput screenings[35]. Another opportunity is the complete and fast characterization of NIR fluorescence nanomaterials such as SWCNTs with quantum defects and potential new synthetic routes[56].

Overall, the NIR-MMS setup provides easy and portable multimodal NIR spectroscopy at a low cost, which makes it useful also for point-of care diagnostics or biosensing in the field for precision agriculture. We anticipate that it closes a technical gap that hindered research groups to start exploring and using the NIR.

**Materials and methods**

For the SCWNT samples 2 mg of SWCNT powder (Sigma-Aldrich, CAS-No 308068-56-6), was suspended in 1 ml of MiliQ water. To make the SWCNTs soluble in aqueous solutions, a modification with biomolecules or surfactants is needed. For the sensing experiments with the neurotransmitter dopamine, the SWCNTs were functionalized with the ssDNA sequence $(GT)_{10}$ and dispersed in phosphate buffered saline (PBS) via tip sonication. Dopamine-hydrochloride (Sigma-Aldrich, MW=189.64 g/mol) was used and diluted from a stock solution of 10 mM dopamine in MiliQ water. The procedure for preparation of ssDNA-SWCNTs can be found in the article by Nißler et al[57]. The total volume for measurements in the wells of a 96 well- plate was 100 µL, consisting of 99 µL of 1 nM $(GT)_{10}$-SWCNT in PBS and 1 µL of dopamine-hydrochloride at varying concentrations.

The diazonium salt used for the quantum defect reaction with 0.5% SDBS dispersed SWCNT was 4-Nitrobenzenediazonium tetra-fluoroborate 97 (Sigma-Aldrich, MW=236.92 g/mol). The defect reaction preparation protocol was based on a paper by Ma et al[49]. The organic IR-1061 dye (Sigma-Aldrich, MW=749.13 g/mol) was dissolved in ethanol, with a final concentration being 100 µg per 1 ml ethanol. The Egyptian blue pigment powder used for exfoliation with the nanosheets was $CaCuSi_4O_{10}$, Kremer Pigmente GmbH & Co. KG as previously described[30].

All fluorescence spectra were recorded in OceanView software using 5 s integration time and 5 scans to average, while absorbance spectra were recorded with 0.5 s integration time and 5 scans to average. The data was processed with a gliding average filter in Python script (numpy.convolve) to smooth intensity fluctuations. It was found that the optimal window size for averaging is n=7 frames, which was applied to the obtained data in all experiments. For further details, we refer the reader to the Support Information document.

## Disclosures

The authors declare no conflict of interest. This work was funded by the Deutsche Forschungsgemeinschaft (DFG, German Research Foundation) under Germany's Excellence Strategy – EXC 2033 – 390677874 – RESOLV. We thank the DFG for funding. This work was supported by the "Center for Solvation Science ZEMOS" funded by the German Federal Ministry of Education and Research BMBF and by the Ministry of Culture and Research of North Rhine-Westphalia. This work was funded by the VW Foundation.

## Supporting Information

The supporting information is available online free of charge. It contains additional spectra, information about data processing and a list of optical components.

# Supporting Information

Multimodal near-infrared spectroscopy for biosensing applications

Krisztian Neutsch[1], Jana Nikolić[1], Asra Mafakheri[3], Jan Stegemann[2], Sebastian Kruss[1,2*]

[1] Department of Chemistry and Biochemistry, Bochum University, Bochum, Germany
[2] Fraunhofer Institute for Microelectronic Circuits and Systems, Duisburg, Germany
[3] Department of Physics, Institute for Advanced Studies in Basic Sciences, Zanjan, Iran

* Email: Sebastian.kruss@rub.de

## List of components

*Table 1.* NIR-MMS setup list of components and their price (without the optional xy stage)

| Component | Price € |
|---|---|
| Stellar-Net SL-1, 350 - 2500 nm visible lamp | 845.12 |
| LED 565nm, Thorlabs M565L3 | 233.24 |
| f = 20.1 mm, Thorlabs SM1U25-A lens | 287.39 |
| FESH0650 Shortpass Filter, Cut-Off 650 nm | 130.72 |
| DMLP805 Longpass Dichroic Mirror, 805 nm Cut-On | 281.84 |
| FELH0850 Longpass Filter, Cut-On Wavelength: 850 nm | 130.72 |
| L2, Olympus LUCPlanFL N, 20x, NA = 0.45 | 1009.31 |
| ZABER ASR100B120B-T3 stage | 5850.87 |
| LA1131 - N-BK7 Plano-Convex Lens, f = 50 mm, Uncoated | 22.42 |
| PL100-2-VIS-NIR fiber bundle | 336.11 |
| Ocean Insight, NIRQUEST+1.7 | 16900 |

# Parameters

Measurement of (GT)$_{10}$ SWCNTs on the setup with varying exposure times and sampling averages shows the influence on the noise behavior. For reduction of noise, we used 5 samples to average. In addition, we used 5 s integration time for fluorescent measurements (as shown in the Figure 2). For absorption measurements, we used 0.5 s integration time.

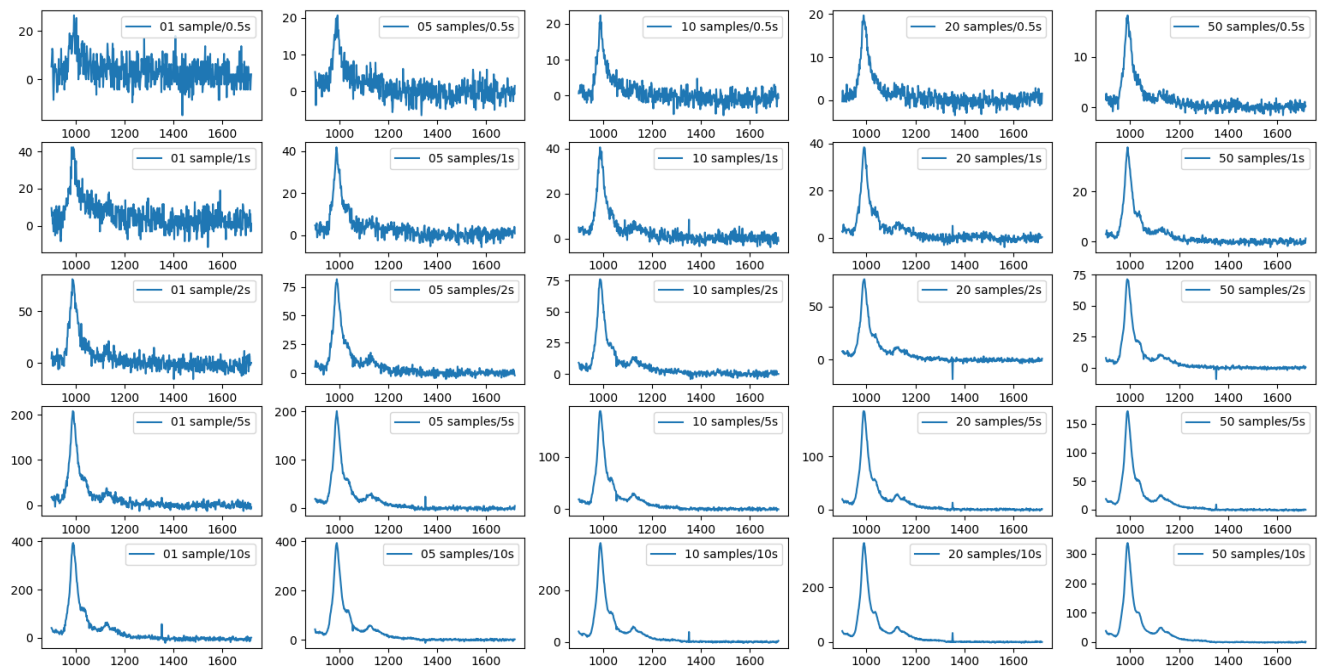

*Figure 1*. Influence of exposure time and sampling (average) on noise of (GT)$_{10}$-SWCNT fluorescence measurements.

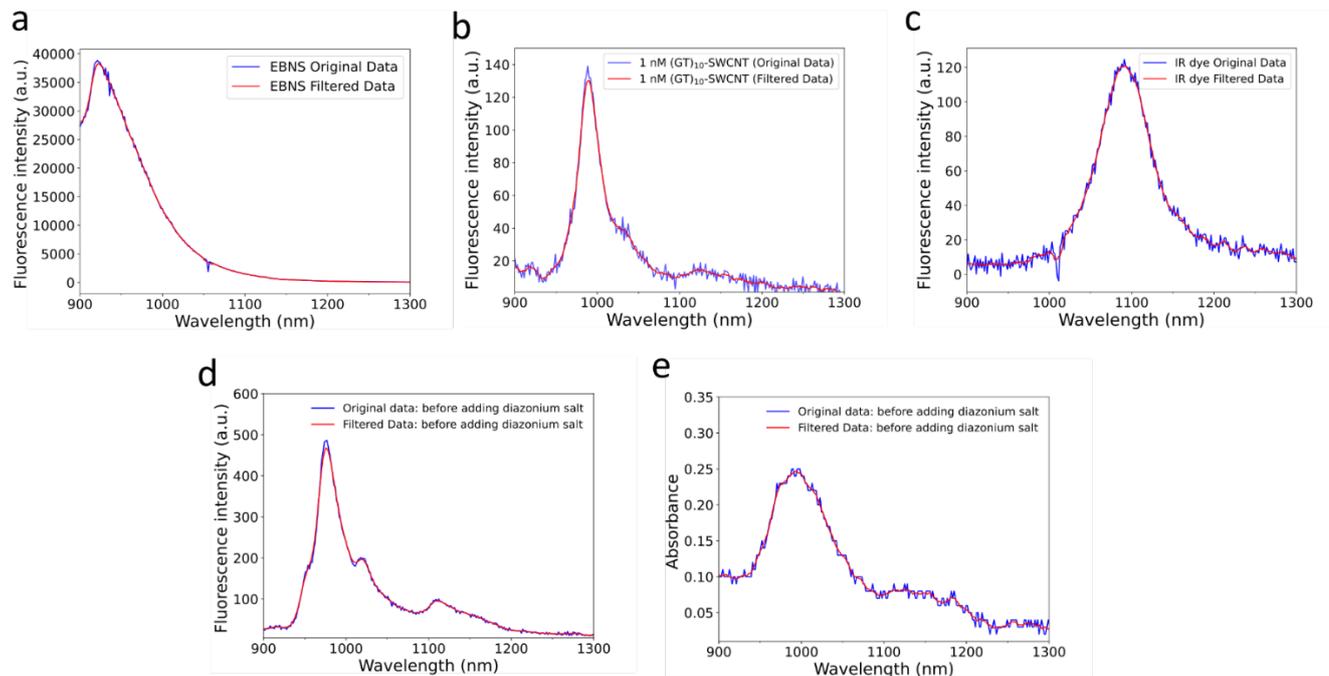

*Figure 2. Original and filtered data of different samples shown for comparison, when using window size n=7 for averaging.* a) Fluorescence spectrum of calcium copper silicate nanosheets (EB-NS) in Mili Q water. b) Fluorescence spectrum of 1 nM $(GT)_{10}$-SWCNTs in PBS. c) Fluorescence spectrum of the NIR fluorescent organic dye IR-1061 (100 µg/ml in ethanol). d) Fluorescence and (e) absorbance spectra of 0.5% SDBS dispersed 5 nM SWCNTs before adding 2 µM diazonium salt. All figures were processed with gliding average filter (numpy.convolve) using n=7 frames for averaging. Comparisons between original unfiltered and filtered data are shown.

## Calibration

Hg-Ar Calibration lamp confirms setup's accuracy and shows a slight wavelength shift of 3 nm in comparison to reference peaks.

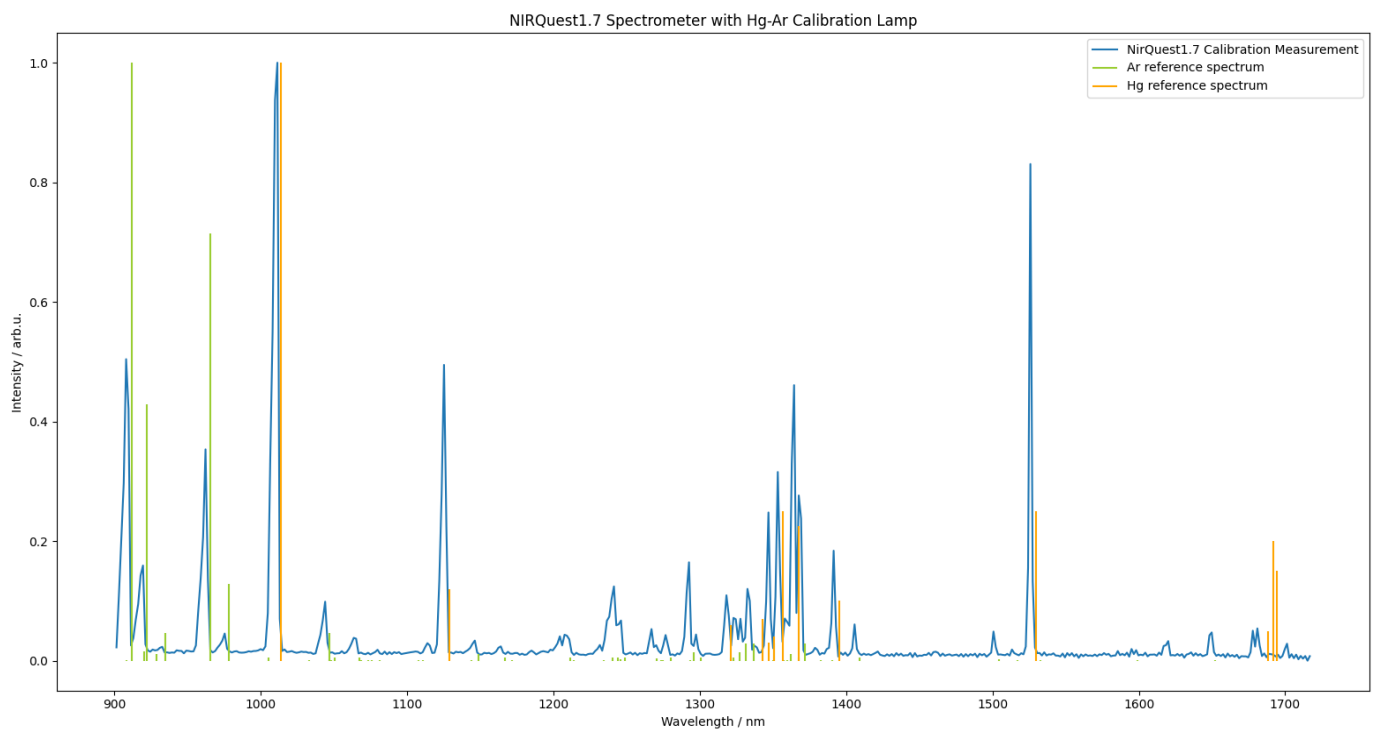

*Figure 3.* NirQuest 1.7 spectrometer output of Ar-Hg calibration lamp emission source (blue) compared to the reference positions of Ar (green) and Hg (orange) atomic lines. The wavelength mismatch between the reference and measured spectrum is approximately 3 nm.